\documentclass[twocolumn,showpacs,preprintnumbers,amsmath,amssymb]{revtex4}
\usepackage{graphicx}
\input{epsf}
\newcommand{\sig}{\mbox{\boldmath{$\sigma$}}}
\begin{document}

\title{Edge excitons in a 2D topological insulator in the magnetic field}

\author{M.V. Entin$^{+,*}$~\thanks{e-mail: entin@isp.nsc.ru}, L.I. Magarill$^{+,*}$, M.M. Mahmoodian$^{+,*}$}
\address{$^{+}$Institute of Semiconductor Physics, Siberian Branch of the Russian Academy of Sciences, Novosibirsk, 630090, Russia \\ $^{*}$ Novosibirsk State University, Novosibirsk, 630090, Russia}

\begin{abstract}
Exciton edge states and the microwave edge exciton absorption of a 2D topological insulator subject to the in-plane magnetic field are studied. The magnetic field forms a narrow gap in electron edge states that allows the existence of edge exciton. The exciton binding energy is found to be much smaller than the energy of a 1D Coulomb state. Phototransitions exist on the exciton states with even numbers, while odd exciton states are dark.
\end{abstract}

\maketitle
\subsection*{Introduction}
The 2D topological insulator (TI) has recently become one of the most popular topics in solid state physics. In this system the inverse electron spectrum of the bulk material produces the gap inside a 2D domain accompanied by the edge states with linear dispersion. The edge spectrum branches are bound to the spin direction so that the propagation direction of the electron with a fixed spin is determined by the edge and spin. There exists voluminous literature, both theoretical and experimental, on the subject (see, e.g., reviews \cite{konig}-\cite{qi} and references therein). The 2D CdTe/HgTe/CdTe quantum well with central layer width $d>$ 6.3 nm \cite{konig2}-\cite{kvon2} is one of the most known representative of 2D TI.

Spin conservation means topological insulation of the edge states: electrons moving along the edge in one direction can not change the direction of motion due to scattering remaining in the same edge. The absence of backscattering leads to non-locality of electron transport \cite{kvon1}-\cite{Roth}. Topological insulation can be violated by spin-flip scattering and interedge transitions \cite{glazm1}-\cite{we}, but these processes are relatively weak.

The edge states in the 2D topological insulator (ESTI) with a linear spectrum can be considered as a 1D analog of graphene. However, unlike graphene, ESTI are freely controlled by the preparation conditions: namely, by the width of CdHgTe quantum well and the content of components.

The single-electron energy spectrum of 2D HgTe layers was studied in \cite{bern}. The edge states  were described in \cite{qi}. In the presence of the in-plane magnetic field the ESTI were considered in \cite{raichev}. It was found that the application of  magnetic field opens the energy gap in the edge states. The excitons in the surface states of a 3D topological insulator, subject to the exchange-induced magnetic action of a contacting magnetic medium were studied in \cite{IonGarate}. At the same time, edge excitons in the 2D TI stay beyond the scope of consideration. The possibility of moving indirect excitons in monolayer graphene was discussed in \cite{makh-ent}. This exciton state is conditioned by the trigonal correction to the linear spectrum; the exciton disappears if its momentum approaches the conic point.

The purpose of the present study is to find the edge exciton states in the presence of magnetic field and to study their influence on ESTI optics. Unlike graphene, ESTI have 1D spectrum and the considerations similar to \cite{makh-ent} are inapplicable to the case. At the same time, in contrast to graphene, a large HgTe g-factor makes the magneto-induced energy gap relatively wide. The finite energy gap gives rise to the exciton states in the ESTI when the system is undoped and unbiased. The Coulomb interaction is  stronger in a 1D system as compared to 3D and 2D. This makes the exciton in ESTI an interesting object of study.

The paper is organized as follows. First, we consider the edge electron states in the presence of magnetic field. Then, the equations for excitons and the formulae for the absorption coefficient will be derived. Next, the numerical solutions will be presented. Finally, we shall discuss the results.

\subsection*{Exciton states}
We consider a HgTe 2D quantum well in the $(x,y)$ plane surrounded by CdS. The edge states  are formed along edge $y=0$ of semi-infinite system $y>0$.

We start from the Hamiltonian of a single electron in the edge state in the presence of external homogeneous magnetic field ${\bf B}=({\bf B}_\|,B_z)$, ${\bf B}_\|=(B_x,B_y)$
\begin{equation}\label{ham011}
   H_e=s\sigma_zp_x+g_{\|}\mu_B\sig\cdot{\bf B}_\|,
\end{equation}
where $\sig=(\sigma_x,\sigma_y,\sigma_z)$ are the Pauli matrices, $p_x$ is the one-dimensional momentum, $\mu_B$ is the Bohr magneton, $g_\parallel$ and $g_\perp$ are in- and out-of-plane components of the $g$-factor. The second term in Eq.~(\ref{ham011}) represents the Zeemann interaction. The Hamiltonian (\ref{ham011}) can be deduced from the 2D Hamiltonian for a CdTe/HgTe/CdTe quantum well (see Appendix).

In the absence of magnetic field at $p_x=0$  Eq.~(\ref{ham011}) gives the double-degenerate state that linearly splits with $p_x$.
The eigenvectors and eigenenergies of the Hamiltonian (\ref{ham011}) are
\begin{eqnarray}
\epsilon_{\pm}=\pm\sqrt{s^2p_x^2+\Delta^2}, \label{en}
\\ \label{fun}\psi_{p_x}^\pm=\frac{1}{\sqrt{\cal L}}e^{ip_xx}
\left(
\begin{array}{c}
 sp_x\pm\sqrt{\Delta^2+s^2p_x^2} \\
 \Delta \\
\end{array}
\right)\frac{1}{N_\pm},
\end{eqnarray}
where $\Delta=|g_{\|}\mu_B{\bf B}_\||=|g_{\|}|\mu_B B, ~~N_\pm=\sqrt{2\sqrt{s^2p_x^2+\Delta^2}(\sqrt{s^2p_x^2+\Delta^2}\pm sp_x)},$ ${\cal L}$ is the  edge length; we set $\hbar=1$. This spectrum has a ''relativistic'' shape with a magneto-induced gap $\Delta$ which can be made arbitrarily narrow.

Using Eq.~(\ref{ham011}), one can write the effective Hamiltonian for an electron-hole pair taking into account the Coulomb interaction $V(x)$:
\begin{eqnarray}
\label{Heh}
 H_{eh}=s(\sigma_{ez} p_{ex}-\sigma_{hz} p_{hx}) + g_{\|}\mu_B(\mbox{\boldmath{$\sigma$}}_e+\mbox{\boldmath{$\sigma$}}_h)\cdot{\bf B}_\|+ \nonumber\\ V(x_e-x_h).
\end{eqnarray}
Here $\mbox{\boldmath{$\sigma_e$}}=\mbox{\boldmath{$\sigma$}}\otimes {\bf I},~~\mbox{\boldmath{$\sigma$}}_h= {\bf I}\otimes \mbox{\boldmath{$\sigma$}}$, {\bf I} is the $2\times 2 $ identity matrix, $x_{e(h)}$ is the electron (hole) coordinate, $p_{e(h)x}=-i\partial_{{e(h)x}}$ is the electron (hole) momentum operator. The Coulomb interaction $V(x_e-x_h)$ here depends on the coordinates along the edge.

Introducing variables of the center of mass and relative motion $X=(x_e+x_h)/2, ~~x=x_e-x_h$, we find from Eq.~(\ref{Heh})
\begin{eqnarray}\label{Heh1}
   H_{eh}=-i\frac{s}{2}(\sigma_{ez} -\sigma_{hz}) \partial_X -is(\sigma_{ez} +\sigma_{hz})\partial_x + \nonumber\\
   g_{\|}\mu_B(\mbox{\boldmath{$\sigma$}}_e+\mbox{\boldmath{$\sigma$}}_h)\cdot{\bf B}_\|+ V(x).
\end{eqnarray}
It is evident enough that the solution of the Schr\"{o}dinger equation will not depend on the magnetic field direction relative to the edge. To be convinced, one should perform the rotation clockwise through an angle $\phi$ ($\phi$ is the angle between ${\bf B}_\|$ and axis $x$) around axis $z$ in spin spaces. Making the corresponding unitary transformation of the Hamiltonian Eq.~(\ref{Heh}) we arrive at
 \begin{eqnarray}\label{Heh2}
   H_{eh}=-i\frac{s}{2}(\sigma_{ez} -\sigma_{hz}) \partial_X -is(\sigma_{ez} +\sigma_{hz})\partial_x+ \nonumber\\
   g_{\|}\mu_B(\sigma_{ex}+\sigma_{hx})B+  V(x).
\end{eqnarray}

The solution of the Schr\"{o}dinger equation is searched in the form of $\Phi(x)e^{iPX}/\sqrt{\cal L}$, where $\Phi(x)$ is the four-component spinor $\Phi=(\Phi_1,\Phi_2,\Phi_3,\Phi_4)$, $P$ is the total momentum. The equation for $\Phi(x)$ reads
\begin{eqnarray}\label{Scheq}
2s(\mp i\partial_x)\Phi_{1,4}=(E-V)\Phi_{1,4}-g_\parallel \mu_BB(\Phi_2+\Phi_3);\nonumber\\
\pm sP\Phi_{2,3}=(E-V)\Phi_{2,3}-g_\parallel \mu_BB(\Phi_1+\Phi_4),
\end{eqnarray}
where $E$ is the energy.

Let us introduce new quantities $F_\pm =\Phi_1 \pm\Phi_4, ~~ G_\pm=\Phi_2 \pm \Phi_3$ satisfying the equations
\begin{eqnarray}\label{ScheqFp}
-4s^2\left(\frac{F_+'}{E-V}\right)'=(E-V)F_+ +\frac{4\Delta^2(E-V)F_+}{s^2P^2-(E-V)^2},
\end{eqnarray}
\begin{eqnarray}\label{FmGpm}
  F_-=-2is\frac{F_+'}{E-V};  \nonumber\\
  G_+=\frac{2\Delta(E-V)F_+}{(E-V)^2-s^2P^2}; \nonumber\\
  G_-=\frac{2\Delta s PF_+}{(E-V)^2-s^2P^2}.
\end{eqnarray}
Note, that Eqs.~(\ref{ScheqFp}) and (\ref{FmGpm}), unlike  Eq.~(\ref{Scheq}), are valid if at least one of $V, ~E, ~P$ is not equal to zero.

Intuitively, due to the one-dimensional character of the edge states, one can use strict 1D e-h Coulomb interaction $V_0(x)=-e^2/\kappa|x|$ ($\kappa$ is the background dielectric constant) for the subsequent consideration. However, due to divergency of the energy of the 1D Coulomb ground level, it is insufficient to use this potential. More accurately, function $V(x)$ is obtained by the integration of 2D Coulomb function $-e^2/(\kappa\mid{\bf r}_e-{\bf r}_h\mid)$ with transversal wave functions $g(y)$ of the edge problem:
\begin{eqnarray}\label{VC}
V(x)=-\frac{e^2}{\kappa}\int \frac{dy_e dy_h}{\sqrt{x^2+(y_e-y_h)^2}}g^2(y_e)g^2(y_h).
\end{eqnarray}
Function $g(y)$ is given by Eq.~(\ref{gy}) in Appendix.

Let $P=0$. Eqs.~(\ref{ScheqFp}) and (\ref{FmGpm}) have a ''non-relativistic'' limit when the characteristic Coulomb and, hence, exciton binding energies are less than $\Delta$. In this case we can modify Eqs.~(\ref{ScheqFp}) and (\ref{FmGpm}):
\begin{eqnarray}\label{non}
\frac{1}{2m}F_+''+(E-2\Delta-V)F_+=0,\\
G_+=F_+,~~~G_-=0,~~~F_-\ll F_+,
\end{eqnarray}
where exciton mass $m=\Delta/(2s^2)$, $E-2\Delta\ll 2\Delta$. Eq.~(\ref{non}) has a simple physical meaning. Let us consider the electron and hole near the bottom (top) of the corresponding bands. Then the case of the Hamiltonian of the pair can be written as $p_e^2/2m_e+p_h^2/2m_h+2\Delta+V(x_e-x_h)$, where $m_h=m_e=2m$. For a pair with the zero total momentum we get the Sch\"odinger equation $(p^2/2m+V(x)+2\Delta-E)\psi=0$ which coincides with Eq.~(\ref{non}).

Generally speaking, the divergency of the energy is limited not only by the above-mentioned finiteness  of the Coulomb potential, but also by the finite forbidden band width $2\Delta$. Hence, the transition to the ''non-relativistic'' case is unjustified. However, it follows from the numerical calculations below that in all the considered region of the parameters, the calculated exciton energies are close to the non-relativistic result.

\subsection*{Microwave absorption}
The edge state absorption can be characterized by the real part of the edge 1D conductivity $\sigma$ at light frequency $\omega$. The absorbing power per edge unit length is $\sigma(\omega) \langle{\cal E}_x^2(t)\rangle=\sigma(\omega) {\cal E}_0^2/2$, where ${\cal E}_x={\cal E}_0 \cos(\omega t)$ is the alternating electric field. If the considered system is a planar grating of HgTe strips, the light absorption through it can be expressed also via the absorptance $a=8\pi\sigma/cd$, where $c$ is the light speed, $d$ is the grating period.

The microwave absorption is determined by the transitions caused by vector potential $A_x(t)=-(c{\cal E}_0/\omega)\sin(\omega t)$. The single-electron Hamiltonian of perturbation is $-es\sigma_zA_x(t)/c$.

First, let us consider the edge free-electron absorption. This absorption is determined by the transition amplitude between the states of the single-electron Hamiltonian.
Using Eqs.~(\ref{en}) and (\ref{fun}) we find for $\sigma(\omega)$
\begin{equation}\label{1d_sigma}
    \sigma(\omega)=\frac{2 e^2s\Delta^2}{\omega^2\sqrt{\omega^2-4\Delta^2}}\theta(\omega^2-4\Delta^2).
\end{equation}

The absorption has a threshold $\omega=2\Delta$ near which it has a singularity $\propto (\omega-2\Delta)^{1/2}$ originating from DOS of the 1D system.

Then, we shall calculate the exciton absorption near threshold $2\Delta-\omega\ll 2\Delta$. In this limit, the absorption coefficient is given by the interband matrix element of the velocity operator $s\sigma_z$ between single-electron wave functions at zero momentum $\psi^{\pm}_0=(1,\pm 1)/\sqrt{2}$ and the scalar exciton wave function $\psi(x)=F_+(x)$ at $P=0$ and $x=0$:
\begin{eqnarray}\label{cond}
&&\sigma(\omega)=\frac{2\pi e^2}{\omega} \sum_n|v_{+,-}|^2|F_+(0)|^2\delta(\omega-E_n)=\nonumber\\  &&\sum_n\gamma_n\delta(\omega-E_n), ~~~\gamma_n=\frac{2\pi e^2s^2}{E_n} |F_+(0)|^2.
\end{eqnarray}
The exciton absorption has delta-functional peaks at the energies of the motionless exciton $E_n=2\Delta-\varepsilon_n$.

The generalization of Eq.~(\ref{cond}) to the case of strong e-h interaction requires accounting for the mixing of electron and hole states with zero momenta. Within Eq.~(\ref{Scheq}) one should find the transition probability between the exciton state and the ''vacuum'' state with zero energy and no potential. The vacuum state can be combined from the zero-energy solutions of Eq.~(\ref{Scheq}) at $V=0$, $P=0$. Due to the degeneracy (double degeneracy of spinors and multiple degeneracy in momentum $p$) there are many zero-energy states. These states are
\begin{eqnarray}\label{zero}\nonumber
F_p^{(1)}=\frac{1}{2{\cal L}\sqrt{\Delta^2+s^2p^2}}(-\Delta,sp,sp,\Delta)e^{ipx},\\
F_p^{(2)}=\frac{1}{2{\cal L}}(0,1,-1,0)e^{ipx}\end{eqnarray}
However, the vacuum corresponds to only one combination of these states:
\begin{eqnarray}\label{vac}
F^{vac}=\sum_p F_p^{(1)},
\end{eqnarray}
which gives the right expression for a transition amplitude to the free electron-hole state, coinciding with the result obtained in the single-electron approach.
With the use of Eqs.(\ref{vac}) and (\ref{zero}) we have
\begin{eqnarray}\label{vacuum}
F^{vac}=\frac{1}{2}(-D(x),-\frac{isD'(x)}{\Delta},-\frac{isD'(x)}{\Delta},D(x)),
\end{eqnarray}
where $D(x)=(\Delta/\pi s)K_0(\Delta |x|/ s) ~~(K_0 $ is the Macdonald function).

As a result we get
\begin{eqnarray}\label{condex}
\sigma(\omega)=\frac{2\pi e^2s^2}{\omega}\sum_n\left|\int_{-\infty}^{\infty}dx D(x)F_+(x)\right|^2\delta(\omega-E_n).
\end{eqnarray}
Near the threshold, Eq.~(\ref{condex}) goes to Eq.~(\ref{cond}). However, the presence of function $D(x)$ expands the applicability of this formula to the case of strong e-h interaction.

\subsection*{Numerical results}
We have done our calculations with the use of the CdTe/HgTe/CdTe system parameters from \cite{qi}. We consider the 7 nm width quantum well for which ${\cal A}=3.645$~eV$\cdot{\AA}$, ${\cal B}=-68.6$ eV$\cdot{\AA}^2$, ${\cal D}=-51.2$ eV$\cdot{\AA}^2$, ${\cal M}=-0.010$ eV. The dielectric constant of CdTe, according to \cite{str}, is $\kappa=10.2$; the g-factor, according to \cite{konig}, is $g_\parallel=-20.5$.

The dependence of the effective interaction potential (see Eq.~(\ref{VC})) on the e-h distance is illustrated in Fig.~1. We see that, at a large $x$, the potential approaches the strict Coulomb potential $V_0(x)=-e^2/\kappa|x|$, but, at a small distance, it is essentially weaker than $V_0(x)$. At the same time, the potential conserves the singularity (weaker than $1/|x|$) at $x\to 0$. This fact is conditioned by the edge wave function singularity. The behavior of the potential at $x\to 0$ is especially important for the exciton ground state which energy diverges in the strict Coulomb case.

We have found the exciton wave functions and energies and optical exciton excitation/recombination probabilities. The results are presented in Figs.~1-3. The ground and excited states energies at $P=0$ {\it versus} magnetic field are shown in Fig.~2. The excited states for a large $n$ approach Coulomb values $me^4/\kappa^2n^2$. They become twins with the distance in pair much less than the distance between twins. This corresponds to the observation made in \cite{makh-ent}: the wave function divergency in the one-dimensional Coulomb problem leads to separation of the x-axis into domains $x>0$ and $x<0$, where the states are independent from each other. The account of the potential finiteness at $x=0$ mixes left and right states and lifts the degeneracy converting the levels into close pairs.

The magnetic field dependence of coefficients $\gamma_n$ describing the exciton absorption is shown in Fig.~3. The odd $n$ states are dark states with zero absorption in the dipole approximation. The even states are optically active. The absorption coefficients grow with $B$ and fall with $n$.

At $B=5$T, the characteristic values of the ground exciton energy and coefficient $\gamma_0$ are $\varepsilon_0=3.528$ meV, $2\Delta-\varepsilon_0=45.386$ meV and $\gamma_0=7.29\cdot$10${}^{-4}$ eV$\cdot\mbox{cm}^2$/s. This value of $\varepsilon_0$ is sufficient for experimental observation at temperatures less than $|\varepsilon_0|/3=13$ K.

To observe the exciton absorption, one should produce the system with multiple edges, for example, a grating with non-overlapping exciton wave functions. These conditions will be satisfied for a grating of strips with width 0.5 $\mu$m and period 2 $\mu$m. If to assume the resonance width of delta-function in Eq.~(\ref{condex}) $\Gamma=5\cdot10^{-2}|\varepsilon_0|$, we obtain the exciton absorptance of the grating $a\approx 0.17$. This value looks quite large to be observed.
\begin{figure}[h]\label{fig1}
\centerline{\epsfxsize=7cm\epsfbox{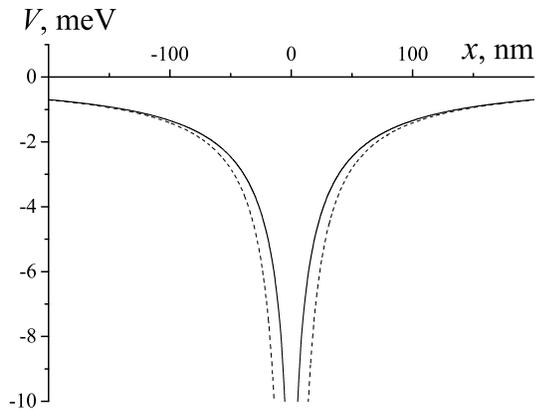}}
\caption{Effective potential of e-h interaction $V(x)$ (solid) and  strict 1D Coulomb potential $V_0(x)$ (dashed).}
\end{figure}
\begin{figure}[h]\label{fig2}
\centerline{\epsfxsize=7cm\epsfbox{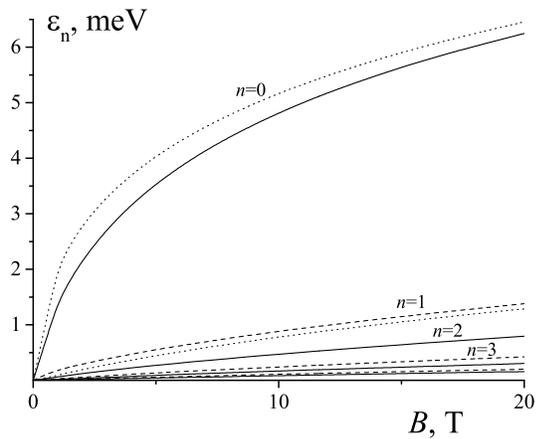}}
\caption{The exciton binding energy at $P=0$ {\it versus} magnetic field. The level number runs from $n=0$ to $n=6$ (downwards). Solid and dashed curves mark even (corresponding to permitted transitions) and odd (forbidden transitions) exciton levels. For comparison the levels $n=0,1$ found in the non-relativistic limit of Eq.~(\ref{non}) are shown by  dotted lines; the difference between curves for other levels is negligible.}
\end{figure}
\begin{figure}[h]\label{fig3}
\centerline{\epsfxsize=7cm\epsfbox{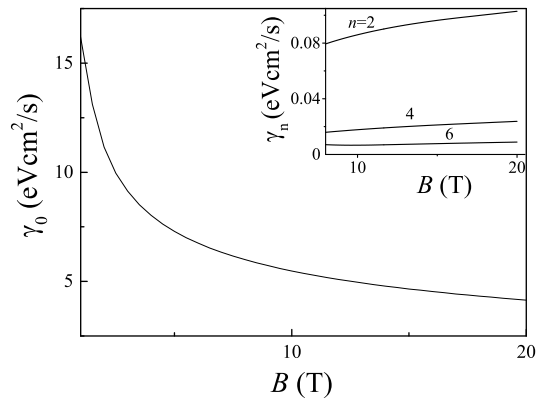}}
\caption{The absorption coefficients for the lowest even exciton states $\gamma_n$ {\it versus} magnetic field.}
\end{figure}
\subsection*{Discussion}
Thus, we have calculated the exciton energy levels and absorption coefficients on exciton transitions for the edge excitons in the 2D topological insulator HgTe with the in-plane magnetic field. The e-h interaction is strong enough to produce edge excitons with the binding energy of the order of $3.5\cdot 10^{-3}$eV and total energy $\sim 4.5\cdot 10^{-2}$eV. The edge exciton states are strongly affected by the smearing of their transversal wave functions, that essentially decreases the exciton energies. At the same time, the mixing of electron and hole states by the Coulomb interaction does not affect them essentially. The excited even and odd exciton states appear in pairs. The absorption on odd excitons is suppressed.

Note, that we have dealt with a two-particle approach only. In fact, the strong enough e-e interaction in the 1D system absolutely reconstructs the many-body system ground state. In this case, many-body effects should be taken into account by the bosonization procedure. The reconstruction of the ground state means accounting for virtual processes with the excitation of multiple e-h pairs. These processes are weak if excitation energy $\Delta$ is large as compared to the Coulomb energy (the validity of this assumption is supposed throughout the present paper). Otherwise, the problem should be considered within the Luttinger liquid approach going beyond the scope of the paper.

Note also, that the edge exciton is the lowest-lying branch of the 2D exciton in a 2D topological insulator. The 2D exciton energy is bound to the 2D  topological insulator gap being essentially larger than the edge exciton energy. One can predict that the edge excitons should play the role of collectors for electron-hole pairs or 2D excitons when the 2D topological insulator is excited. This process would essentially suppress 2D photoconductivity, but simultaneously let the edge excitons to serve as low-energy photon emitters.
\subsection*{Acknowledgements}
This research was supported by RFBR grants No 13-0212148 and No 14-02-00593.
\subsection*{Appendix}
The edge states with a linear spectrum appear due to splitting of double degenerate states at zero momentum. Without the magnetic field, this fact is described by Hamiltonian $H_e=s\sigma_zp_x$. In the presence of the magnetic field, $H_e$ should be complemented by the linear in magnetic field terms. To obtain the magnetic part of edge-state Hamiltonian (\ref{ham010}), we shall use the 2D Hamiltonian proposed in \cite{bern2} and complete it by the Zeemann term of the form (see, e.g., \cite{maciejko}):
\begin{eqnarray}
\label{Z2D}
\mu_B\left(\begin{array}{cccc} g_{E\perp}B_z & 0 & g_\parallel B_-  & 0 \\ 0 & g_{H\perp}B_z & 0 & 0 \\ g_\parallel B_+ & 0 & -g_{E\perp}B_z & 0 \\ 0 & 0 & 0 & -g_{H\perp}B_z\end{array}\right),
\end{eqnarray}
and the vector-potential of magnetic field ${\bf A}$ by replacement of electron momentum ${\bf p}\to {\bf p}-e{\bf A}/c$. Here $B_{\pm}=B_x \pm iB_y$, $g_\parallel,~g_{E\perp},~ g_{H\perp}$ are in-plane and out-of-plane $g$-factors of a 2D topological insulator.

In the first order of the magnetic field, the 1D Hamiltonian is given by the projection of Eq.~(\ref{Z2D}) onto the transversal wave functions for a single edge  without the magnetic field (\cite{zhou}, \cite{we}). These functions, corresponding to near-degenerate states, are

\begin{eqnarray}\label{psiy}
\psi_+(y)= \frac{g(y)}{\sqrt{1+\eta^2)}}
\left(
\begin{array}{c} 1\\ \eta\\ 0 \\ 0
\end{array}
\right),
\nonumber ~~~
\psi_-(y)= \frac{g(y)}{\sqrt{1+\eta^2}}
\left(
\begin{array}{c} 0 \\ 0\\ 1\\ \eta
\end{array}
\right),
\nonumber
\end{eqnarray}
where $\eta=\sqrt{({\cal B+D})/({\cal B-D})},$
\begin{eqnarray}\label{gy}
  g(y)=(e^{-\lambda_1 y}-e^{-\lambda_2 y})\frac{\sqrt{2 \lambda_1\lambda_2(\lambda_1+\lambda_2)}}{\lambda_1-\lambda_2}, \\
\lambda_{1,2} = \frac{\cal A}{2\sqrt{{\cal B}^2-{\cal D}^2}}\pm \sqrt{\frac{{\cal A}^2}{4({\cal B}^2-{\cal D}^2)}-\frac{{\cal M}}{\cal B}}, \nonumber
\end{eqnarray}
${\cal A}, {\cal B}, {\cal D}, {\cal M}$ are the HgTe layer parameters (they are determined by the material parameters and the quantum well width).
As a result, we have
\begin{equation}\label{ham010}
   H_e=s\sigma_z\left(p_x-\frac{e}{c}\overline{A}_x\right)+\mu_B\left[g_\parallel\sig\cdot{\bf B}_\|+g_\perp\sigma_zB_z\right].
\end{equation}
Here $\overline{A}_x$ is the mean value of the vector-potential averaged with the transversal edge-state wave functions $g(y)$. The second term in Eq.~(\ref{ham010}) represents the Zeemann part of the Hamiltonian, where $g_\perp=(g_{H\perp}+g_{E\perp})/2+(g_{H\perp}-g_{E\perp}){\cal D}/(2{\cal B})$.

In the case of a constant homogeneous magnetic field $A_x$ and $B_z$ can be excluded from the Hamiltonian (\ref{ham011}) by a non-essential shift of momentum  $p_x\to p_x+e A_x/c-\mu_Bg_\perp B_z/s$. Finally, we get Eq.~(\ref{ham011}).

Note that, in the chosen approximation neglecting the bulk induced anisotropy, the $B_z$-induced gap is absent; besides, the states have the isotropic in-plane g-factor. A more general case (see discussion in \cite{nature}) can be considered similarly.

\end{document}